# Scintillator cubes for 3D neutrino detector SuperFGD

**S Fedotov, A Dergacheva, A Filik, M Khabibullin, A Khotjantsev, Yu Kudenko, O Mineev, N Yershov**

Institute for Nuclear Research of the Russian Academy of Sciences, 60-letiya Oktyabrya prospekt 7a, Moscow 117312, Russia

e-mail: fedotov@inr.ru

**Abstract**. SuperFGD, a highly granular scintillator detector, is under construction to reduce systematic uncertainties in the T2K experiment in order to improve the sensitivity to CP-violation in neutrino oscillations. SuperFGD will be comprised of about $2\times10^6$ small ($10\times10\times10$ mm$^3$) optically isolated polystyrene based plastic scintillator cubes with three orthogonal holes 1.5 mm in diameter. The readout of scintillating light from each cube is provided by three wavelength shifting fibers inserted into the three holes and coupled to MPPC micropixel photosensors. The cubes are covered with a white chemical reflector for optical isolation. The technology of making these cubes, their mechanical properties, their main characteristics obtained during tests with cosmic muons and at the CERN beamline, and the results of the temperature tests are presented in this paper.

## 1. SuperFGD Detector

The T2K experiment [1] is an accelerator neutrino oscillation experiment carried out by the T2K International Collaboration in Japan. The main goals of this experiment are the study of neutrino oscillation parameters and search for CP-violation in the lepton sector. The main element of the experimental setup is the far detector Super-Kamiokande (SK). This is a large water Cherenkov detector, located 295 kilometers away from J-PARC (Japan Proton Accelerator Research Complex) serving as a neutrino source. The second element of the experimental setup is the near detector system. The system consists of two elements: the INGRID detector located on the beam axis, and the ND280 detector located at an angle of 2.5 degrees to the beam axis. In the T2K experiment the technology of a neutrino beam, shifted from the primary proton beam, is used. This technology makes it possible to get a narrow-band neutrino beam.

In 2017 the T2K collaboration launched the Near Detector Upgrade project [2]. The upgrade is targeted at reducing systematic errors in the T2K search for CP-violation in the lepton sector. The motivation for the ND280 upgrade is driven by the following reasons:

- the current tracker, 2 FGDs and 3 TPCs, detects charged particles from neutrino interactions mostly in forward or backward directions, while the far detector, SK, has a 4π-acceptance;
- to improve the precision of reconstructing the neutrino energy in the charged-current quasi elastic interactions (CCQE) in the near detector, in addition to the 1-track events (with one lepton detected) it would be useful to detect 2-track events (lepton + proton), which requires reducing the proton detection threshold. In the current configuration, this threshold is quite high, and the neutrino energy is reconstructed from 1-track events only;

- to improve the precision of reconstructing the antineutrino energy it is very important to detect neutrons produced in the CCQE interactions of the antineutrinos;
- the sensitivity of the T2K and Hyper-Kamiokande [3] experiments to CP-violation requires reducing the systematic errors, which can be achieved by accurate measurements of the neutrino interaction cross sections.

To address these problems, the concept of a new fine-grained neutrino detector SuperFGD has been proposed (Figure 1) [4]. The main element of the detector is a small scintillation cube of 1 centimeter size. Each cube is covered with a reflector and has three orthogonal holes of 1.5 mm diameter for WLS fibers. The detector consists of 192×182×56 (~2 mln) cubes. In this configuration the total mass of the detector is about 2 tons. Signals from each cube will be read out via three orthogonal 1.0 mm Kuraray Y11 fibers, one end of each fiber will be viewed by the photosensor S13360-1325PE Hamamatsu MPPC.

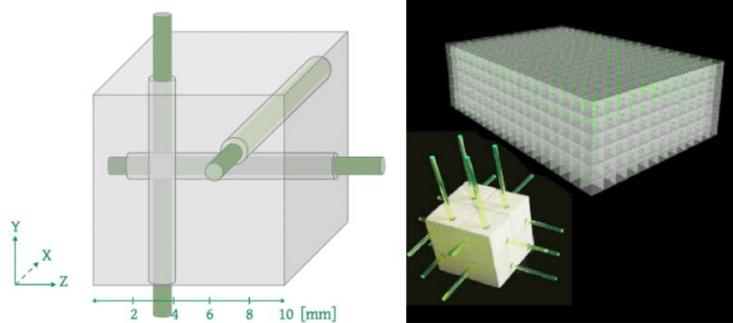

**Figure 1.** Left: a scheme of a scintillation cube for the SuperFGD detector. Right: a concept of the SuperFGD detector.

As the results of MC-simulation show, such detector configuration will help lower the proton momentum registration threshold from the current level of 400-500 MeV/c down to 200-300 MeV/c. Also, the SuperFGD detector will significantly increase the efficiency of detecting particles at angles close to vertical direction [5]. The SuperFGD detector is able to detect muon and neutron from the antineutrino interaction: muon gives a prompt scintillation, while the neutron can produce an energetic proton, so the time-of-flight method can be used to reconstruct the neutron energy [6]. The SuperFGD neutron detection efficiency is relatively high if this method is used. It is expected that this will improve the antineutrino energy resolution to 7% relative to 15%, which is expected using the conventional neutrino energy reconstruction techniques.

## 2. Production and geometric properties of the cubes

The cubes are produced by Uniplast Co. in Vladimir (Russia) using injection molding. The scintillator is made of polystyrene with 1.5% of paraterphenyl and 0.01% of POPOP. After molding, the size of the clear cubes' side is 10.026 mm. Next, the cubes are etched with a chemical agent to form a white microporous reflector layer. The fluctuations of the size of the cubes, after this step, are less than 30 μm. At the final stage of the production, each cube is placed on a jig designed to hold a cube in place during the drilling of the three orthogonal through holes of 1.5 mm diameter. The precision of the hole position relative to the side of the cube is ~40 μm, relative to the angle of the cube is ~50 μm. The precision of the hole position relative to the hole on the opposite side of the cube is ~40 μm. The achieved geometric precision allows the complete assembly of the detector with 1 mm wavelength shifting fibers of 1 mm diameter. The cubes production for the SuperFGD detector has been completed in January 2021. About 95% of the cubes have been selected according to the geometric parameters for the assembly in the detector. The control of the geometric parameters was carried out during the primary assembly of the detector on the fishing lines, which form the 3D structure of the detector with a specified position for each cube. Full-size prototypes were used to test the fishing line method. One of the prototypes consisted of 5 full-size planes of 192×184 cubes. Another real-size prototype consisted of 56 planes of 192×15 cubes. The tests demonstrated that the fishing line can be smoothly replaced with the WLS fiber. All tested fishing lines were easily replaced without any problems.

The pitch between the holes is also crucial, because the holes should be made in the mechanical box in advance, before the assembly. The pitch in the horizontal plane was measured using two full-size prototypes. The first consisted of the five planes (192×184 cubes) which were assembled first, the second consisted of the five full-size planes which were assembled last. The horizontal pitch was equal to 10.30 mm. The vertical pitch is more difficult to measure because the lower layers of the cubes experience pressure from the mass of the upper layers. It was measured using different prototypes, but more precise results were obtained during the test of the prototype shown in Figure 2. This prototype consisted of 56 planes of 15×18 cubes each. It was assembled on fishing lines pulled through the three components. At the first step of the test, the additional load (14 kg) was installed on the top of the prototype for faster prototype compression over time. When the compression stopped increasing, the additional load was removed. The vertical pitch was measured as well (10.28 mm). The fairly strong dependence of the prototype size on temperature can also be observed. This is important to consider when choosing materials for the mechanical box.

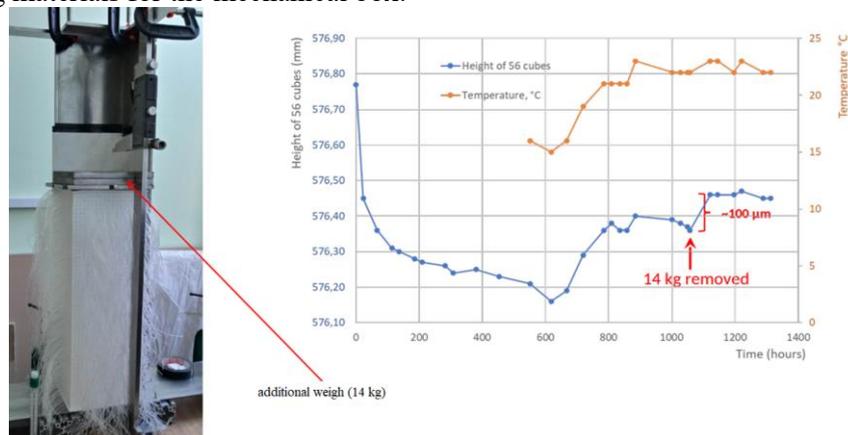

**Figure 2.** The experimental setup and the result of measuring the vertical pitch between the holes of the cubes.

## 3. Control of scintillation properties of the cubes

The scintillation properties of randomly selected cubes from each shipment were regularly checked with cosmic muons. In these tests scintillation light from 24 cubes was transmitted through 8 optical WLS fibers (35 cm Kuraray Y11) to eight photosensors (Hamamatsu MPPC S13081-050C) attached to one end of each fiber (the other end was polished). The test results are shown in Figure 3: for two years of tests over 2000 cubes were checked, the average light yield (l.y.) is ~37 p.e./MIP, and no cubes with l.y. <30 p.e./MIP were found.

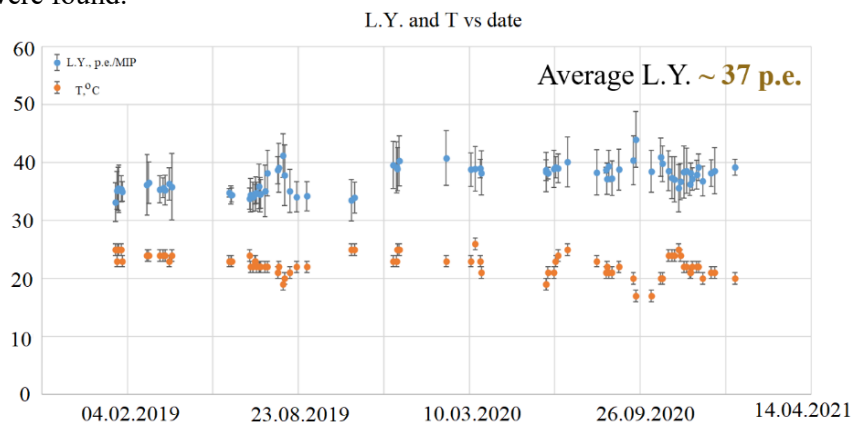

**Figure 3.** The blue points are average light yield from 24 cubes (p.e./MIP), the orange points are the temperature during tests (degrees Celsius).

## 4. Beam tests of the two SuperFGD detector prototypes

Tests of the two SuperFGD detector prototypes were carried out at the Proton Synchrotron (PS) of CERN. The first prototype of SuperFGD was tested at CERN in autumn 2017 [7, 8], and the second prototype was tested in summer 2018 [9]. The results of measuring the main parameters of the two prototypes are in a good agreement with each other (taking into account different electronics, fiber length, etc.). Typical l.y. from a single cube per fiber was measured to be 40 p.e./MIP, two readout fibers produced l.y. of ~80 p.e./MIP. Time resolution per single fiber was 0.95 ns, two fibers provided the resolution of 0.65-0.71 ns. Optical crosstalk through a side of a cube was found to be around 3%.

In this article, we present some details of the tests of the first SuperFGD prototype consisted of 125 (5×5×5) cubes. The scintillations were transmitted by the WLS fibers (Kuraray Y11, Ø1 mm, 1.3 m length) to the photosensors (Hamamatsu MPPC 12571-025C, 1×1 mm$^2$ active area) and read out by the ~5 GHz digitizer (12 channels). Two 3×3×10 mm$^3$ counters located upstream and downstream of the prototype were used for triggering the beam events. In these tests, the cube was scanned with a beam at a pitch of 2 mm. As seen in Figure 4. the horizontal fibers show a stable l.y. (red points), while the vertical fibers demonstrate fluctuations of l.y. within ~20% (blue points), which is related to the beam positions close to the vertical fibers.

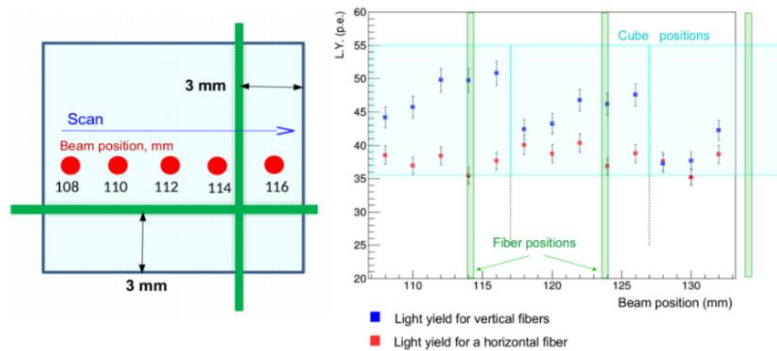

**Figure 4.** Left: Beam scan across a single cube. Right: Light yields for horizontal and vertical fibers.

## 5. Temperature test of the cubes and the larger scintillation cubes

6 groups of cubes were used for temperature tests: the reference one (at room temperature) and 5 groups, being heated up to 50º, 60º, 70º, 80º and 90 ºC, respectively, for 8 hours. The results are shown in Figure 5. The decrease of l.y. level in these tests is related to the chemical properties of the scintillator. After reaching a temperature of 80-90º, an irreversible chemical reaction begins, and the scintillation properties begin to deteriorate. It can be observed visually on the cubes without a reflector. The cubes become opaque after heating to 70º. This is even more noticeable when it comes to the cubes heated to 90º (the rightmost cubes in the photo in Figure 5). As expected, it is dangerous to heat the detector to a temperature >70º.

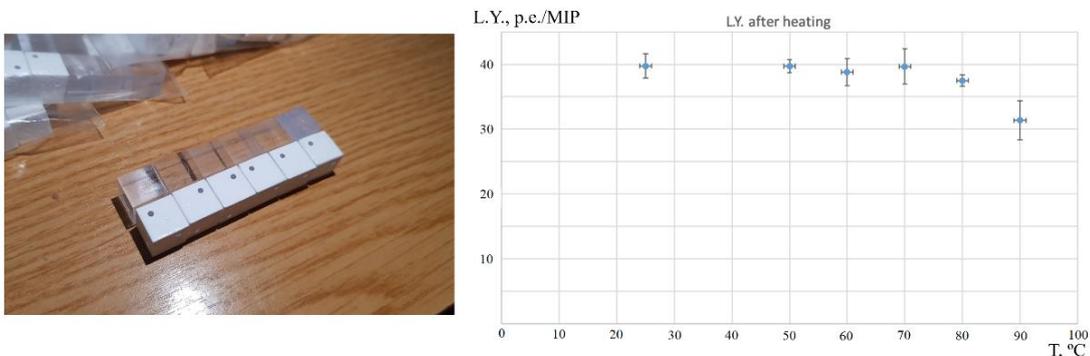

**Figure 5.** Left: a photo of the cubes from the temperature tests (from left to right: the reference group, the cubes heated to 50º, 60º, 70º, 80º and 90º, respectively). Right: l.y. of the cubes after heating for 8 hours at the corresponding temperature.

Large (1.5 and 2.0 cm) cubes can be used in other detectors with similar requirements, for example, at the near detector of the DUNE experiment [10]. At Uniplast Co. large cubes were cut out from the extruded slabs (the cubes for SuperFGD were produced using injection molding). Tests of the large cubes with cosmic muons showed that the energy-normalized light yield from the large cubes is the same as that of the 1-cm cubes used in the SuperFGD detector: ~11 p.e./MeV.

## 6. Conclusion

The cubes production for the SuperFGD detector has been completed. The achieved geometric precision (the fluctuations of cubes' size <30 μm, the fluctuations of the hole positions <50 μm) allows the complete assembly of the detector with wavelength shifting fibers of 1 mm diameter. About 95% of the cubes were selected according to the geometric parameters for the assembly in the detector. The pitch of the holes in the SuperFGD detector mechanical box is the following: the horizontal pitch is 10.30 mm and the vertical pitch is 10.28 mm. For two years of tests over 2000 cubes were checked, the average l.y. is ~37 p.e./MIP, and no cubes with l.y. <30 p.e./MIP were found. The time resolution of the cube per single fiber was 0.95 ns, two fibers provided the resolution of 0.65-0.71 ns. The optical crosstalk through a side of the cube was found to be ~3%. The degradation of the scintillation properties is observed when the cubes are heated to a temperature higher than 70 ºC. Energy-normalized l.y. from large cubes (1.5 and 2.0 cm) is the same as the l.y. from the cubes used in the SuperFGD detector (1.0 cm cubes). The average energy-normalized l.y. is ~11 p.e./MeV.

## 7. Acknowledgements

This work is partly supported by the RSF grant №19-12-00325 and the JSPS-RFBR grant №20-52-50010.